# Influence of mass contrast in alloy phonon scattering

Takuma Shiga[1], Takuma Hori[1], and Junichiro Shiomi[1,2,*]

[1]*Department of Mechanical Engineering, The University of Tokyo, Bunkyo, Tokyo 113-8656, Japan*
[2]*Japan Science and Technology Agency, PRESTO, Saitama 332-0012, Japan*

We have investigated the effect of mass contrast on alloy phonon scattering in mass-substituted Lennard-Jones crystals. By calculating the mass-difference phonon scattering rate using a modal analysis method based on molecular dynamics, we have identified the applicability and limits of the widely-used mass-difference perturbation model in terms of magnitude and sign of the mass difference. The result of a phonon-mode-dependent analysis reveals that the critical phonon frequency, above which the mass-difference perturbation theory fails, decreases with the magnitude of the mass difference independently of its sign. This gives rise to a critical mass contrast, above which the mass-difference perturbation model noticeably underestimates the lattice thermal conductivity.

[*] E-mail: shiomi@photon.t.u-tokyo.ac.jp

# 1. Introduction

After a long history of study in lattice (phonon) heat conduction in crystals, accurate calculations of microscopic phonon transport properties and lattice thermal conductivity have become accessible by the recent development of methodologies to extract anharmonic interatomic force constants from first-principles.[1,2] The calculations typically employ the anharmonic lattice dynamics[1-4] or molecular dynamics (MD)[3] and have been successfully applied to various single crystals with a wide range of lattice thermal conductivity from ~1 to ~$10^3$ Wm$^{-1}$K$^{-1}$.[1-9] The accurate anharmonic interatomic force constants have been also used to calculate lattice thermal conductivity of alloyed crystals or solid solutions by adopting the perturbation models[6,8,10] or the Green-Kubo formula.[3,11] Retaining accuracy in such calculations is particularly important for thermoelectric applications, where alloying has been widely employed to reduce lattice thermal conductivity to improve the material figure of merit.[12]

While alloying gives rise to local mass and force field contrasts, most of the works so far have focused on the lattice thermal conductivity reduction due to the former. In the view of phonon gas transport, this would correspond to the mass-difference phonon scattering. The simplification of the alloy effect has relevance in thermoelectric materials alloyed with homologous element, although there are certainly materials where the local force field contrast matters.[11] Klemens[13] has theoretically derived the quartic frequency dependence of mass-difference phonon scattering rates ($\gamma_{ms} \propto \omega^4$, where $\omega$ is the angular frequency) in analogy to the Rayleigh scattering. Abeles[14] was the first to incorporate the Klemens' formula into the Callaway model,[15] and qualitatively reproduced the reduction of lattice thermal conductivity of silicon-germanium alloys.

Later, Tamura[16] has rigorously derived the mass-difference phonon scattering rates based on the perturbation theory (referred here as *Tamura model*). For uniform monoatomic system, the scattering rate of phonon with the wavevector (**q**) and branch (*s*) due to mass-difference, considering lowest-order perturbation, is expressed as

$$\gamma_{\text{Tamura}}(\mathbf{q}s) = \frac{\pi}{6N}\omega^2(\mathbf{q}s)g_2\sum_{\mathbf{q}'s'}\delta\{\omega(\mathbf{q}s)-\omega(\mathbf{q}'s')\}, \quad (1)$$

where $N$ denotes the total number of primitive unit cells in the system. The constant $g_2=f(1-m_i/m)^2$ is a measure of the strength of the mass-difference phonon scattering, where $f$ and $m_i$ are the concentration and mass of the substituted atoms, respectively, and $m$ is the average atomic mass of the system.

The Tamura model has been implemented into the anharmonic lattice dynamics with the first-principles-based interatomic force constants, and calculations of lattice thermal conductivity have been reported for silicon-germanium,[10] lead chalcogenides,[6] and magnesium-silicide-based materials.[8] Although some of the calculations[8,10] have successfully reproduced the lattice thermal conductivity in alloys even at 50 % concentration, application of the model to high concentration and/or large mass contrast requires care since the model is based on the lowest-order perturbation theory. Larkin and McGaughey[17] have calculated the mass-difference phonon scattering rates of mass-substituted argon and silicon crystals by MD and lattice dynamics for various alloy concentrations and found an evident discrepancy with the Tamura model in the high frequency regime even with the lowest alloy concentration (5 %). It was suggested that the discrepancy is due to lack of higher-order terms in the mass-difference perturbation. Interestingly, the above discrepancy was found to have a minor impact on thermal conductivity in the case of the mass-substituted silicon crystal since the heat is dominantly conducted by low-frequency phonons.

In addition to the concentration dependence, another important parameter in the mass-difference perturbation is the mass ratio ($m_i/m$). Since alloys or solid solutions with large mass contrast are popular class of materials when aiming to reduce the thermal conductivity such as in thermoelectric applications,[18] it is important to gain understanding in applicability and limit of the Tamura model in terms of the mass contrast. To this end, we have investigated the influence of the mass ratio on the mass-difference phonon scattering rates by using the MD-based modal analysis method. By taking the mass-substituted Lennard-Jones (argon) alloy crystals as a representative case, the analysis was performed for various mass ratios at a fixed low alloy concentration (2 %).

## 2. Method

We have used the normal mode projection (NMP) method[19, 20] to calculate the phonon scattering rate. The method has been applied to various pure crystals and alloys.[21-27] In the NMP method, a normal mode coordinate and its conjugate momentum are computed by projecting atomic trajectories obtained by MD simulations onto the eigenvector (**e**).

Using the displacement vectors (**u**) obtained from MD simulations, the normal mode coordinate [$Q(\mathbf{q}s;t)$] with given wavevector (**q**) and phonon branch ($s$) at time $t$ is written as

$$Q(\mathbf{q}s;t) = \sum_k \sqrt{\frac{M_k}{N}} \mathbf{e}_k^* \cdot \mathbf{u}_k(t) \exp(i\mathbf{q} \cdot \mathbf{R}_k), \qquad (2)$$

where $N$ is the total number of primitive unit cells. $M_k$ and $\mathbf{R}_k$ denote the mass and equilibrium position of the $k$th atom, respectively. The vector $\mathbf{e}_k$ is a component of the eigenvector at the $k$th atom, and the asterisk denotes the complex conjugate. The normal mode momentum [$P(\mathbf{q}s;t)$] is computed similarly by simply replacing the displacement vector in Eq. (2) with the velocity vector.

Under the quasi-harmonic approximation,[28, 29] the normal mode energy [$E(\mathbf{q}s;t)$] can be written as

$$E(\mathbf{q}s;t) = \frac{1}{2}P^*(\mathbf{q}s;t)P(\mathbf{q}s;t) + \frac{1}{2}\omega^2(\mathbf{q}s)Q^*(\mathbf{q}s;t)Q(\mathbf{q}s;t), \qquad (3)$$

where $\omega(\mathbf{q}s)$ is the angular frequency of $\mathbf{q}s$ phonon. Finally, the phonon scattering rate [$\gamma(\mathbf{q}s)$], an inverse of phonon relaxation time, is obtained from the temporal decay of the autocorrelation coefficient:

$$\frac{\langle \tilde{E}(\mathbf{q}s;t)\tilde{E}(\mathbf{q}s;0) \rangle}{\langle \tilde{E}(\mathbf{q}s;0)\tilde{E}(\mathbf{q}s;0) \rangle} = \exp\{-\gamma(\mathbf{q}s)t\}. \qquad (4)$$

Here, $\tilde{E}$ denotes the deviation from the time-averaged normal mode energy.

In this work, for the sake of simplicity and for comparison with the previous reports,[17, 20, 30] we have chosen mass-substituted Lennard-Jones alloyed crystals. Lennard-Jones (12-6) potential, which has been widely used to describe the van der Waals interaction, is given by

$$V(r_{ij}) = 4\varepsilon_{\mathrm{LJ}}\left\{\left(\frac{\sigma_{\mathrm{LJ}}}{r_{ij}}\right)^{12} - \left(\frac{\sigma_{\mathrm{LJ}}}{r_{ij}}\right)^{6}\right\}, \qquad (5)$$

where $r_{ij}$ is the distance between $i$th and $j$th atoms. The parameters of argon, $\varepsilon_{\mathrm{LJ}}=1.67\times10^{-21}$ J, $\sigma_{\mathrm{LJ}}=3.4\times10^{-10}$ m, and $m=39.948$ amu were adopted.[20, 29] The cut-off radius of the interaction was set to $2.5\sigma_{\mathrm{LJ}}$.

After equilibration, constant energy MD simulations (i.e., microcanonical ensemble) were performed at 20 K for 8.0 ns with the time step of 4.0 fs, and normal mode energies in the entire first Brillouin zone were calculated using Eq. (3). For each mass difference, 30 MD simulations with different initial conditions were performed to ensemble-average $\chi(\mathbf{q}s)$. A 6×6×6 conventional cubic supercell with 864 atoms was placed in a fully periodic cell. The lattice constant was set to 5.315 Å, which is the pressure-corrected value at 20 K.[20]

We prepared the mass-substituted Lennard-Jones alloyed crystals by randomly substituting atoms in the supercell. The mass ratio of substituted to host atoms ranges from $m_i/m$=0.2 to 5.0. For each mass ratio, the simulations were performed for three different configurations for configurational averaging. The alloy concentration was fixed at 2 %.

When applying the NMP method to the alloyed crystals, we have used eigenstates of the pure crystal following the assumption in the Tamura model. Although the mass-substitution breaks the crystal invariance in the supercell and reduces the wavevector space to the Γ point, we can intuitively expect the eigenstates to be similar to those of the pure crystal when the alloy concentration is low. Larkin and McGaughey[17] have shown the validity of the assumption for the mass ratio $m_i/m$=3.0 and concentration ranging from 5 to 50 % in terms of the structure factor.[17,31,32] While this may justify the use of the pure crystal eigenstates for the current alloyed crystals, it is certainly questionable if the argument is valid for larger or smaller mass ratio. To this end, we evaluated the dependence of the structure factor on the mass ratio.

According to Larkin and McGaughey[17], the structure factor for eigenstate with wavevector **q** is defined as follows:

$$S_\text{L}(\mathbf{q},\omega) = \sum_s \left| \sum_j \hat{\mathbf{q}} \cdot \boldsymbol{e}_j(\mathbf{k}=\mathbf{0},s)\exp(i\mathbf{q}\cdot\mathbf{R}_j) \right|^2 \delta\{\omega - \omega(\mathbf{k}=\mathbf{0},s)\}, \qquad (6)$$

$$S_\text{T}(\mathbf{q},\omega) = \sum_s \left| \sum_j \hat{\mathbf{q}} \times \boldsymbol{e}_j(\mathbf{k}=\mathbf{0},s)\exp(i\mathbf{q}\cdot\mathbf{R}_j) \right|^2 \delta\{\omega - \omega(\mathbf{k}=\mathbf{0},s)\}, \qquad (7)$$

where subscripts L and T denote longitudinal and transverse components, respectively. Here, $\hat{\mathbf{q}}$ is the unit vector of $\mathbf{q}$.

Figures 1(a) and 1(b) respectively show the structure factors for $m_i/m=0.2$ (the smallest case) and 5.0 (the largest case) for $\mathbf{q}$ vectors along the Γ-X symmetry line. The red and blue lines indicate the longitudinal and transverse components of the structure factor, respectively. Since the calculated $S_{L(T)}$ has a distribution over the frequency, we have calculated the peak centers through the first-order moment, $\omega_{c,L(T)}(\mathbf{q}) = \int_{-\infty}^{\infty} \omega \tilde{S}_{L(T)}(\mathbf{q},\omega)d\omega$, where $\tilde{S}_{L(T)}$ is the normalized structure factor. Figures 1(c) and 1(d) show the deviation from the frequency of the pure crystal mode ($|\omega_c-\omega_{pure}|/\omega_{pure}$) as a function of the mass ratio. The figures show that the deviation is at most 6 %, suggesting the validity of using the pure crystal eigenstates for all the mass-ratio cases studied in this work. It is noted that the 6 % deviation here corresponds to less than 1 % in terms of the criterion used by Larkin and McGaughey.[17]

## 3. Results and discussions

### 3.1 Phonon scattering rates

Firstly, we have calculated phonon scattering rates due to the intrinsic phonon-phonon scattering ($\gamma_{p\text{-}p}$) by applying the NMP method to the pure crystal. Figure 2 shows the frequency dependence of $\gamma_{p\text{-}p}$ at 20 K. Note that all phonon branches are acoustic since the primitive unit cell of the pure system consists of only one atom. As seen in Fig. 2, $\gamma_{p\text{-}p}$ is widely distributed from 0.0 to 0.7 THz, which is consistent with the work reported by McGaughey and Kaviany.[20] The intrinsic scattering rate $\gamma_{p\text{-}p}$ is proportional to the squared frequency ($\gamma_{p\text{-}p} \propto \omega^2$) in the low frequency regime, which is in agreement with the Klemens' formula.[33]

We have next applied the NMP method to the mass-substituted Lennard-Jones alloyed crystals. Figure 3 shows the frequency dependences of the total phonon scattering rates ($\gamma_c$) for six different mass ratios ($m_i/m$=0.2, 0.5, 1.5, 2.0, 3.0, and 5.0) with 2 % alloy concentration. As shown in Fig. 3, the profiles of $\gamma_c$ are similar to those of $\gamma_{p-p}$ since $\gamma_c$ includes the intrinsic phonon-phonon scattering. On the other hand, there is a distinct difference between the profiles of $\gamma_c$ and $\gamma_{p-p}$ due to the contribution from the mass-difference phonon scattering. This enables us to calculate phonon scattering rates due to the mass-difference ($\gamma_{ms}$) using the Matthiessen's rule[28, 29] [$\gamma_{ms}(\mathbf{q}s)=\gamma_c(\mathbf{q}s)-\gamma_{p-p}(\mathbf{q}s)$].

Figure 4 shows the frequency dependences of $\gamma_{ms}$ for $m_i/m$=0.2, 0.5, 1.5, 2.0, 3.0, and 5.0. The figures show that, for all mass ratios, $\gamma_{ms}$ has the quartic frequency dependence ($\gamma_{ms} \propto \omega^4$) in low frequency regime, which agrees with the Klemens[13] and Tamura[14] models. On the other hand, the overall frequency dependence distinctively varies with the mass ratio, which disagrees with the models that predict the self-similar frequency dependence for variation of the mass ratio [Eq. (1)]. Another noteworthy point is that, although the magnitude of mass-difference for the cases of mass ratio $m_i/m$=0.5 and 1.5 are the same, their $\gamma_{ms}$ clearly differs from each other in the high frequency regime, where $\gamma_{ms}$ increases and decreases with frequency, respectively. Note that, as seen in Eq. (1), the mass-ratio dependence of $\gamma_{Tamura}$ is symmetric around $m_i/m$=1.0.

Figure 5 shows the direct comparisons of averaged $\gamma_{ms}$ and $\gamma_{Tamura}$ for $m_i/m$=0.2, 0.5, 1.5, 2.0, 3.0, and 5.0 as a function of frequency. Here, the moving average in the frequency space is applied to the raw data of $\gamma_{ms}$ to reduce the influence of statistical noise. Although the large number of MD samplings in this work reduces the statistical uncertainties of $\gamma_c$ and $\gamma_{p-p}$ to less than a few percent, since $\gamma_{ms}$ is calculated from their difference (Matthiessen's rule), the statistical uncertainty of $\gamma_{ms}$ grows as the mass contrast approaches 1 (i.e., $\gamma_{ms}$ becomes small). With the moving average, as shown in Fig. 5, the spread of $\gamma_{ms}$ becomes small enough to be compared with $\gamma_{Tamura}$. As has been reported by Larkin and McGaughey[17], $\gamma_{Tamura}$ reasonably agrees with $\gamma_{ms}$ in the low frequency regime, but the discrepancy is evident in the high frequency regime. We here find that the critical frequency above which the discrepancy becomes evident significantly depends on the mass ratio. The discrepancy can be quantified in terms of the difference $\Delta\gamma(\omega)=|\gamma_{ms}(\omega)-\gamma_{Tamura}(\omega)|$. By introducing a threshold value of scattering rate ($\gamma_{thr}$), we define the critical phonon frequency $\omega_{cf}$ as the minimum frequency that satisfies $\Delta\gamma(\omega) > \gamma_{thr}$. Here, we take $\gamma_{thr}$ from the characteristic intrinsic phonon-phonon scattering rate, obtained through the simplest expression of the lattice thermal conductivity based on the phonon gas model $\kappa_{lat}=(1/3)cv^2/\gamma_{thr}$[34], where $c$ and $v$ denote the volumic specific heat and speed of sound, respectively. By setting $v=1145$ ms$^{-1}$, we obtain $\gamma_{thr}=0.043$ THz under the classical approximation. Figure 6 shows that $\omega_{cf}$ decreases as $m_i/m$ increases or decreases from 1.0, indicating that the applicable frequency range of the Tamura model shrinks with increasing mass contrast. Although the current analysis is qualitative due to the ambiguity in the choice of $\gamma_{thr}$, it clearly shows the general trend how Tamura model gradually fails in the frequency domain by increasing the mass contract, which becomes important in understanding how the discrepancy influences the lattice thermal conductivity calculation as will be discussed in the next section.

3.2 Lattice thermal conductivity

Let us now investigate the impact of the above discrepancy in the phonon scattering rates on the lattice thermal conductivity. Having the phonon scattering rates, the lattice thermal conductivity ($\kappa_{lat}$) can be calculated based on the phonon gas model:[34]

$$\kappa_{lat} = \frac{1}{3}\sum_{\mathbf{q}s} c(\mathbf{q}s)v_g^2(\mathbf{q}s)\gamma^{-1}(\mathbf{q}s), \qquad (8)$$

where $c(\mathbf{q}s)$ and $v_g(\mathbf{q}s)$ respectively denote the volumetric specific heat and group velocity of $\mathbf{q}s$ phonon, which were calculated using the lattice dynamics method.[28] The phonon scattering rate ($\gamma$) was obtained in two different ways: one is to directly use the calculated $\gamma_c$ (Fig. 2), and the other is to use the Tamura model through the Matthiessen's rule[28, 29] [$\gamma(\mathbf{q}s)=\gamma_{p\text{-}p}(\mathbf{q}s)+\gamma_{\text{Tamura}}(\mathbf{q}s)$]. Here, we will refer the lattice thermal conductivity calculated from the former and the latter as $\kappa_{lat}^c$ and $\kappa_{lat}^{\text{Tamura}}$, respectively.

In addition to the phonon gas model, the lattice thermal conductivity was calculated based on the Green-Kubo formula with the heat flux vector ($\mathbf{J}_Q$) obtained by the MD simulations:[35]

$$\kappa_{lat}^{\text{GK}} = \frac{\Omega}{3k_B T^2} \int_0^\infty \langle \mathbf{J}_Q(t) \cdot \mathbf{J}_Q(0) \rangle dt, \qquad (9)$$

where $\Omega$, $k_B$, and $T$ denote the volume of supercell, the Boltzmann constant, and temperature, respectively. $\mathbf{J}_Q$ was sampled every 400 fs during the MD simulations, and $\kappa_{lat}^{\text{GK}}$ was averaged over 30 MD simulations with different initial conditions.

Figure 7 shows the dependence of lattice thermal conductivities ($\kappa_{lat}^c$, $\kappa_{lat}^{\text{Tamura}}$, and $\kappa_{lat}^{\text{GK}}$) on the mass ratio at 20 K. Each lattice thermal conductivity is normalized by its value for the pure system ($m_i/m=1.0$). All the profiles generally follow the same trend, where the thermal conductivity decreases as $m_i/m$ deviates from 1.0. The profiles collapse on top of each other in 0.5<$m_i/m$<1.5, which indicates the validity of the Tamura model to express lattice thermal conductivity in this regime. On the other hand, as the mass contrast increases, $\kappa_{\text{lat}}^{\text{Tamura}}$ gradually deviates from $\kappa_{\text{lat}}^{\text{c}}$ and $\kappa_{\text{lat}}^{\text{GK}}$. The above discussion on the critical frequency suggests that the discrepancy in lattice thermal conductivity increases with the mass contrast because $\omega_{\text{cf}}$ decreases, and the discrepancy becomes significant when $\omega_{\text{cf}}$ falls in the range of frequency with dominant contribution to lattice thermal conductivity. While the mechanism of the reduction in $\omega_{\text{cf}}$ is likely to be the lack of higher order terms in the mass-difference perturbation, it is worth noting that the Tamura model appears to underestimate the lattice thermal conductivity. The underestimation is consistent with the report of Larkin and McGaughey.[17] It is not clear at this point how exactly the higher-order terms of the mass-difference perturbation should influence the scattering rate and thus lattice thermal conductivity. Further investigation of such aspect remains to be our future task.

In this work, we show that whether the Tamura model is applicable to lattice thermal conductivity calculation of mass-substituted alloy depends significantly on the mass contrast in addition to the concentration. This can be described in terms of the two competing characteristic frequencies: (1) the maximum frequency of phonons with noticeable contribution to thermal conductivity and (2) the critical frequency above which the Tamura model fails to describe the mass-difference phonon scattering. Although the current calculations were performed for argon crystal, the qualitative discussion should be applicable to influence of mass contrast in the alloy crystals in general including polyatomic systems. Despite some success in reproducing the lattice thermal conductivity of alloyed crystals[18], the current work shows that there certainly should be a limit to the applicability of the model in terms of the mass contrast depending on the lattice thermal conductivity spectra (frequency-dependent thermal conductivity). It is interesting to note that the applicability may not simply decrease with increasing concentration since higher concentration typically shrinks the thermal conductivity spectra towards lower frequency, which may make the Tamura model more applicable as far as the lattice thermal conductivity is concerned. Similarly, even for a material where the failure of the Tamura model at high frequency regime does not influence the obtained total lattice thermal conductivity, the influence could manifest when the material is nanostructured, where the interface is expected to dominantly scatter low-frequency phonons, resulting in the relative increase in the contribution of the high-frequency phonons to lattice thermal conductivity.[36]

## 4. Conclusions

We have investigated the influence of mass ratio ($m_i/m$) to mass-difference phonon scattering rate by performing the modal analysis on molecular dynamics (MD) of mass-substituted Lennard-Jones (argon) crystals, and evaluated the applicability and limit of the Tamura model. The phonon-mode analysis identifies the critical phonon frequency, above which the mass-difference scattering rate predicted by the Tamura model deviates from that calculated directly by MD. This can result in significant underestimation of lattice thermal conductivity when the critical phonon frequency is small with respect to the maximum frequency of phonons with noticeable contribution to thermal conductivity. The mass-ratio dependence study finds that the critical phonon frequency decreases as the mass ratio increases or decreases from 1.0, which narrows the applicable frequency regime of the Tamura model. Such a dependence of the applicability of the Tamura model on the mass ratio is expected to be found also in other materials including polyatomic ones.

**Acknowledgements**

This work was partially supported by Research Fellowships of the Japan Society for the Promotion of Science for Young Scientists, Japan Science and Technology Agency PRESTO, the Thermal & Electric Energy Technology Foundation, and KAKENHI 23760178.

**Figure captions**

**Fig. 1.** (Color online) Structure factors for (a) $m_i/m=0.2$ and (b) 5.0 calculated using Eqs. (5) and (6) for **q** vectors along the Γ-X symmetry line. Red and blue lines denote the longitudinal and transverse components of the structure factor, respectively. $q_X$ is the length of Γ-X symmetry line. Mass-ratio dependences of the deviation ($|\omega_c-\omega_{pure}|/\omega_{pure}$) in (c) longitudinal and (d) transverse components.

**Fig. 2.** (Color online) Frequency dependence phonon scattering rates due to intrinsic phonon-phonon scattering ($\gamma_{p-p}$) of pure Lennard-Jones crystal at 20 K. Red marks denote the results by McGaughey and Kaviany.[20] The dashed solid line in the inset figure indicates the Klemens' formula ($\gamma_{p-p} \propto \omega^2$).[33]

**Fig. 3.** (Color online) Frequency dependent phonon scattering rates of mass-different Lennard-Jones alloyed crystals ($\gamma_c$) with 2 % alloy concentration at 20 K for six different mass ratios: (a) $m_i/m=0.2$ and 0.5, (b) $m_i/m=1.5$ and 2.0, (c) $m_i/m=3.0$ and 5.0. The intrinsic phonon-phonon scattering rate $\gamma_{p-p}$ is also plotted for comparison.

**Fig. 4.** (Color online) Frequency dependent phonon scattering rates ($\gamma_{ms}$) due to the mass-difference scattering of six different mass ratios ($m_i/m$) calculated through the Matthiessen's rule using $\gamma_c$ and $\gamma_{p-p}$: (a) $m_i/m=0.2$ and 0.5, (b) $m_i/m=1.5, 2.0, 3.0$, and 5.0. Dashed lines in the insets (a) and (b) show the frequency dependence, $\gamma_{ms} \propto \omega^4$.

**Fig. 5.** (Color online) Direct comparisons between averaged $\gamma_{ms}$ and $\gamma_{Tamura}$ for six different mass ratios.

**Fig. 6.** (Color online) Critical phonon frequency ($\omega_{cf}$) as a function of mass ratio ($m_i/m$).

**Fig. 7.** (Color online) Lattice thermal conductivities ($\kappa_{lat}^{c}$, $\kappa_{lat}^{Tamura}$, and $\kappa_{lat}^{GK}$) normalized by those of pure crystal as a function of mass ratio ($m_i/m$).

**Fig. 1.**

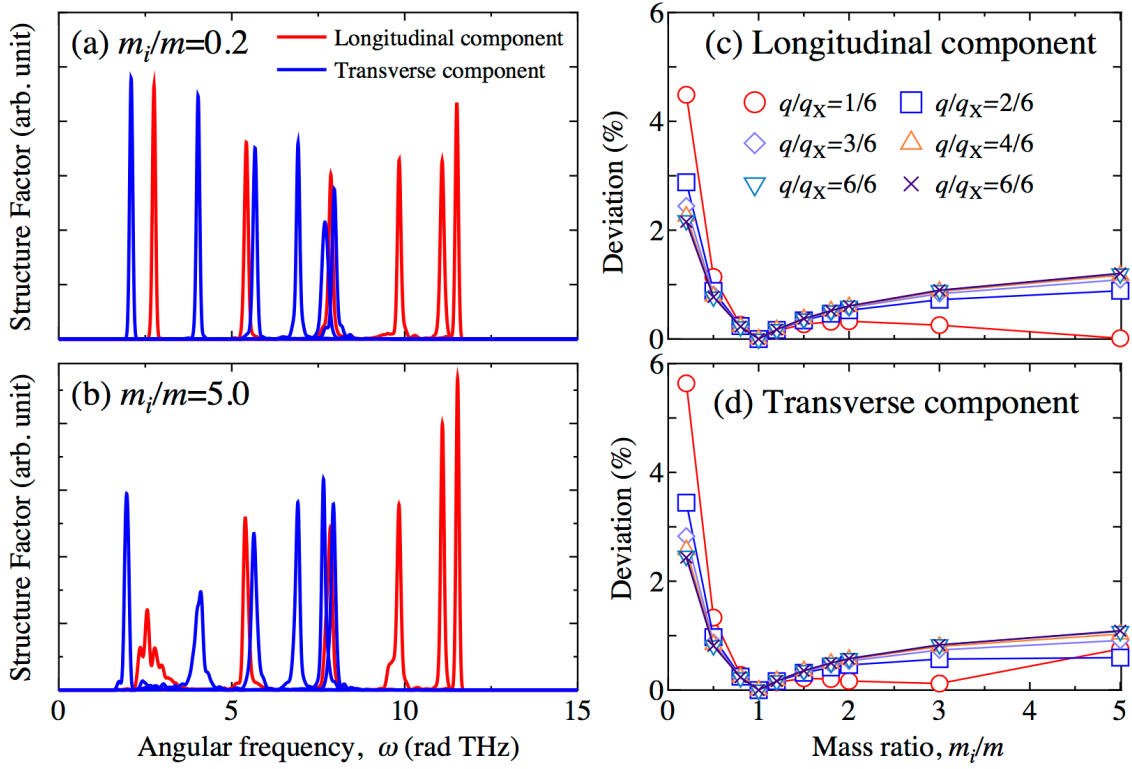

**Fig. 2.**

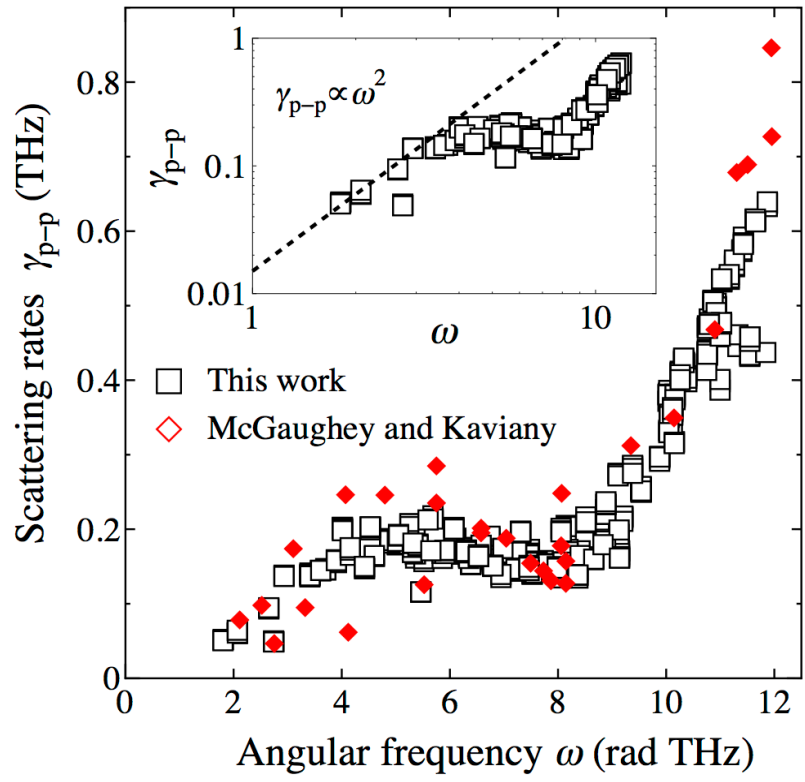

**Fig. 3.**

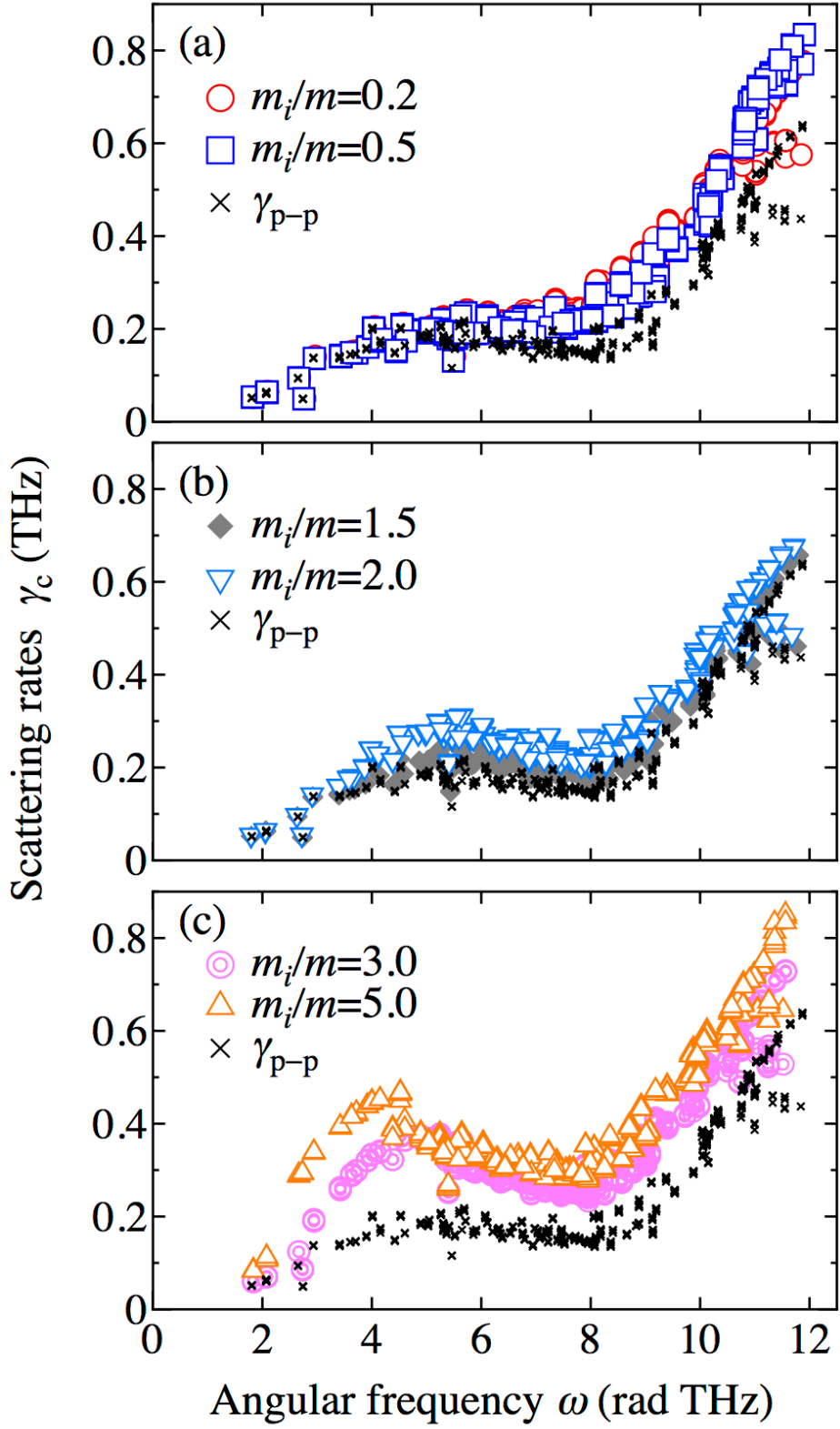

**Fig. 4.**

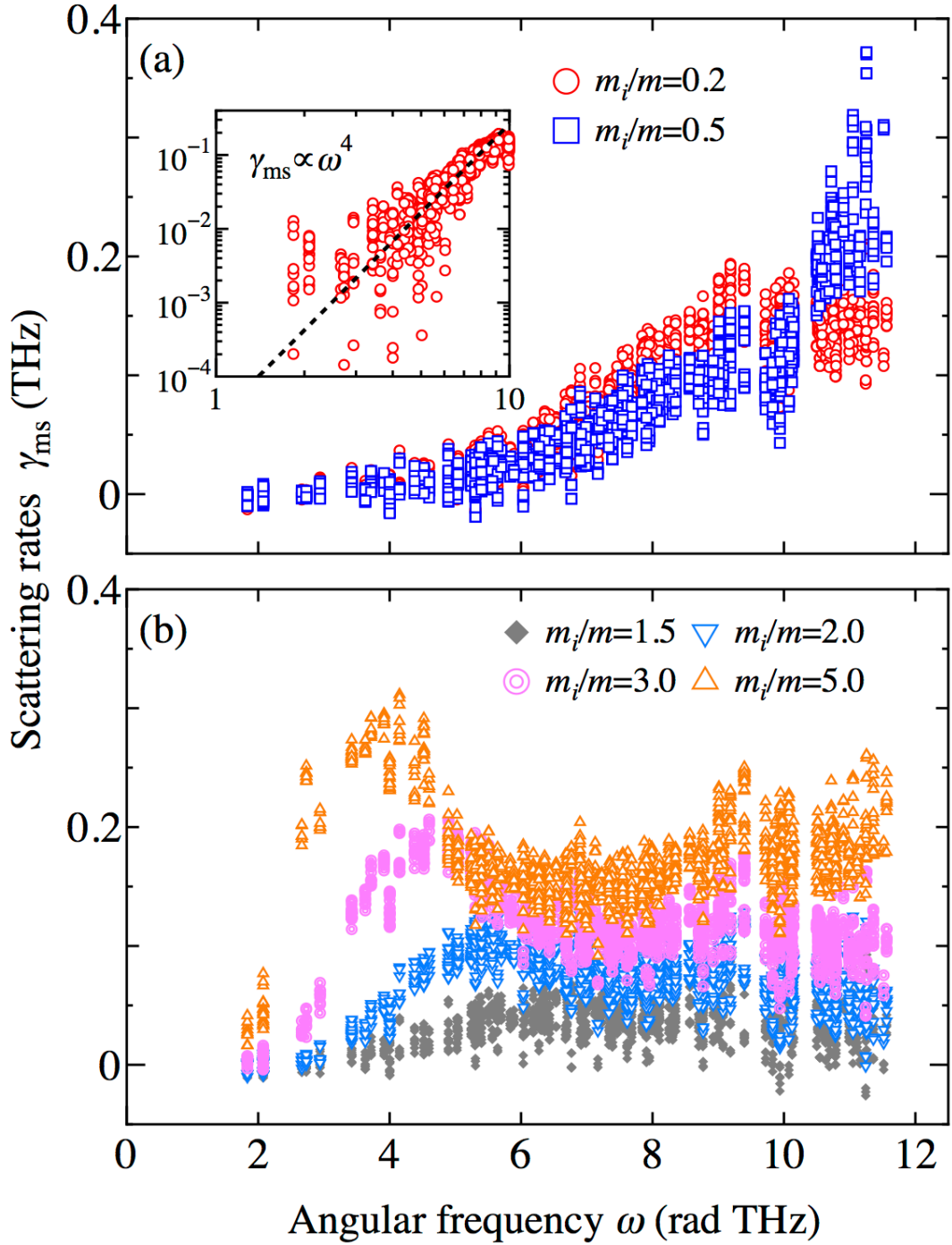
2222

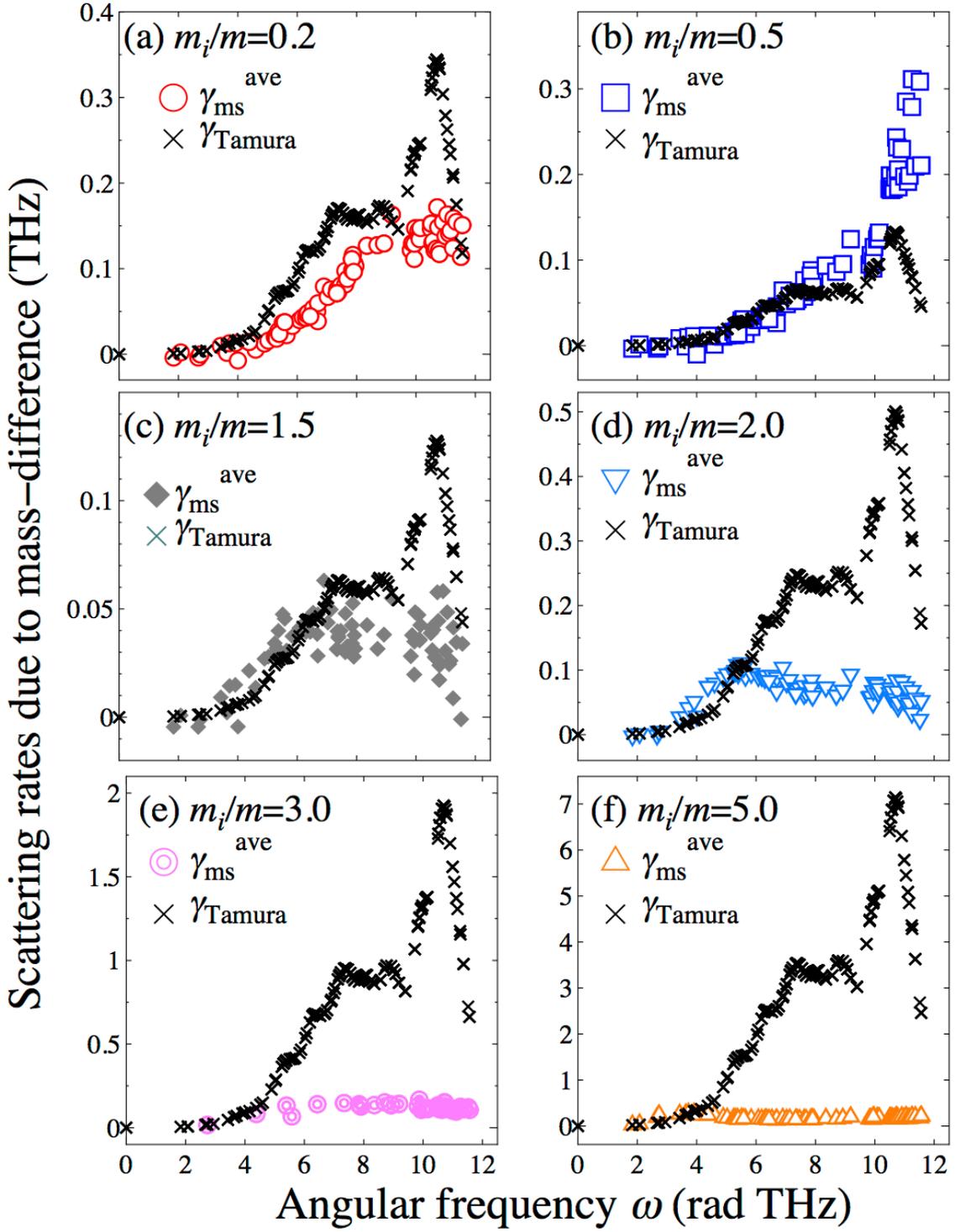

**Fig. 6.**

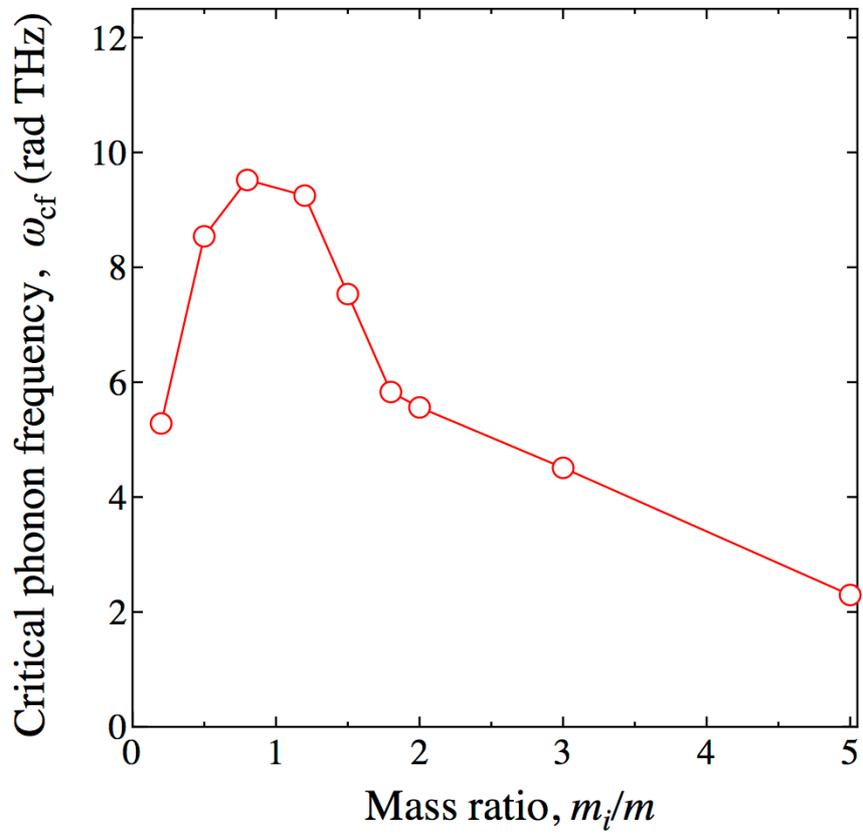

**Fig. 7.**

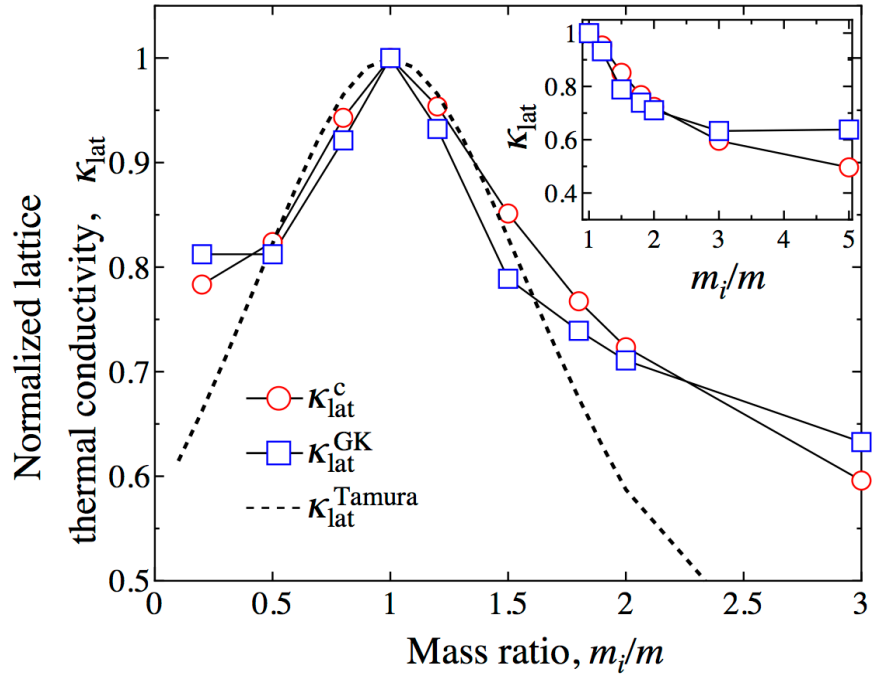